\documentclass[twocolumn,prb,aps,showpacs,,superscriptaddress]{revtex4-1}

\usepackage{amsmath,amssymb,amsfonts}
\usepackage{color}
\usepackage[colorlinks,breaklinks,bookmarks=true,citecolor=blue,linkcolor=black,urlcolor=blue]{hyperref}
\definecolor{darkred}{rgb}{0.7,0.0,0}

\usepackage{graphicx}
\usepackage{multirow}

\begin{document}

\preprint{APS/123-QED}

\title{Crucial role of out-of-plane Sb-$p$ orbitals in Van Hove singularity formation and electronic correlation for superconducting Kagome metal CsV$_3$Sb$_5$}

\author{Min Yong Jeong}
 \affiliation{Department of Physics, Korea Advanced Institute of Science and Technology (KAIST), Daejeon 34141, Korea}

\author{Hyeok-Jun Yang}
 \affiliation{Department of Physics, Korea Advanced Institute of Science and Technology (KAIST), Daejeon 34141, Korea}

\author{Hee Seung Kim}
\affiliation{Department of Physics, Korea Advanced Institute of Science and Technology (KAIST), Daejeon 34141, Korea}

\author{Yong Baek Kim}
\email{ybkim@physics.utoronto.ca}
\affiliation{Department of Physics, University of Toronto, Toronto, Ontario M5S 1A7, Canada}

\author{SungBin Lee}
\email{sungbin@kaist.ac.kr}
\affiliation{Department of Physics, Korea Advanced Institute of Science and Technology (KAIST), Daejeon 34141, Korea}

\author{Myung Joon Han}
\email{mj.han@kaist.ac.kr}
\affiliation{Department of Physics, Korea Advanced Institute of Science and Technology (KAIST), Daejeon 34141, Korea}

\date{\today}

\maketitle
{\bf
First-principles density functional theory calculations are performed to understand the  electronic structure and  interaction parameters for recently discovered superconducting Kagome metal CsV$_3$Sb$_5$. A systematic analysis of the tight-binding parameters based on maximally localized Wannier function method demonstrates that the out-of-plane Sb$^{\rm out}$-$p$ orbital is a key element in complete description of the three Van Hove singularity structures known in this material at $M$ point near the Fermi level. Further, the correlation strengths are also largely determined by Sb$^{\rm out}$-$p$ states. Based on constrained random phase approximation, we find that  on-site and inter-site interaction parameter are both significantly affected by the screening effect of Sb$^{\rm out}$-$p$ orbitals. As the role of this previously unnoticed orbital state can be tuned or controlled by out-of-plane lattice parameters, we examine the electronic structure and particularly the evolution of Van Hove singularity points as a function of strain and pressure, which can serve as useful knobs to control the material properties. 
    } 



\label{sec:intro}

The recently discovered Kagome metal family $A$V$_3$Sb$_5$ ($A=$K, Rb, Cs) has gained a lot of  attention by displaying various types of ordered phases intertwined with one another, such as charge density wave (CDW) \cite{ortiz_new_PhysRevMaterials.3.094407_first_Synthesize,shumiya_intrinsic_2021,ortiz_fermi_2021,li_observation_2021,ortiz_fermi_2021,ratcliff_coherent_2021,zhao_cascade_2021,kang_microscopic_2022,hu_coexistence_2022,jiang_unconventional_2021,song_orbital_2021,wang_electronic_PhysRevB.104.075148,liang_three-dimensional_2021,chen_roton_2021}
and superconductivity \cite{ortiz_topological_PhysRevLett.125.247002_firstSC,chen_double_2021,yu_unusual_2021,zhang_pressure-induced_2021,du_pressure-induced_2021,chao_mu1_s-wave_2021,duan_nodeless_2021,xu_multiband_stmSCgap_PhysRevLett.127.187004,xiang_twofold_2021,wang_proximity-induced_2020,shunli_ni_anisotropic_2021,chen_roton_2021,liang_three-dimensional_2021,wang_electronic_PhysRevB.104.075148,zhao_nodal_2021}. Several different CDW orders have been observed and/or suggested, including the one with broken time-reversal symmetry \cite{shumiya_intrinsic_2021,jiang_unconventional_2021,song_orbital_2021,ortiz_fermi_2021,li_observation_2021,ortiz_fermi_2021,ratcliff_coherent_2021,kang_microscopic_2022,hu_coexistence_2022,liang_three-dimensional_2021,zhao_cascade_2021,wang_electronic_PhysRevB.104.075148,chen_roton_2021,yang_giant_2020_anomalousHall,Yu_Concurrence_PhysRevB.104.L041103_anomalousHall_Cs,denner_analysis_2021,feng_chiral_2021,feng_lowenergy_PhysRevB.104.165136,lin_complex_PhysRevB.104.045122_2021,yu_evidence_2021_muSR,mielke_iii_time-reversal_2021_muSR}. Whereas some important experimental features remain controversial \cite{neupert_charge_2021,jiang_Kagome_2021,wang_electronic_PhysRevB.104.075148,song_orbital_2021,xu_multiband_stmSCgap_PhysRevLett.127.187004,wang_proximity-induced_2020,xiang_twofold_2021,shunli_ni_anisotropic_2021,jiang_unconventional_2021,Tan_CDW_DFT_PhysRevLett.127.046401}, theoretical studies have been actively conducted in order to provide a plausible understanding of their microscopic origins and the inter-relationships between different orders, stimulating further explorations \cite{chen_double_2021,yu_unusual_2021,zhang_pressure-induced_2021,du_pressure-induced_2021,chao_mu1_s-wave_2021,duan_nodeless_2021,xu_multiband_stmSCgap_PhysRevLett.127.187004,xiang_twofold_2021,wang_proximity-induced_2020,shunli_ni_anisotropic_2021,chen_roton_2021,liang_three-dimensional_2021,wang_electronic_PhysRevB.104.075148,zhao_nodal_2021}.


In spite of key insights from previous studies \cite{feng_chiral_2021,feng_lowenergy_PhysRevB.104.165136,denner_analysis_2021,lin_complex_PhysRevB.104.045122_2021,park_electronic_2021,gu_gapless_2021_singleband_MLWF_model,wang_competing_PhysRevB.87.115135_2013,Kiesel_unconvenitonal_PhysRevLett.110.126405_2013,Kiesel_sublattice_PhysRevB.86.121105_2012,yu_chiral_superconducting_PhysRevB.85.144402_2012,wu_nature_2021}, this intriguing material system is still far from being clearly understood. Importantly, the constructed phase diagrams are markedly different from each other, even in the simplest case of single orbital (typically taking $d_{xy}$ orbital) Kagome model calculations at Van Hove filling \cite{wang_competing_PhysRevB.87.115135_2013,Kiesel_unconvenitonal_PhysRevLett.110.126405_2013,Kiesel_sublattice_PhysRevB.86.121105_2012,yu_chiral_superconducting_PhysRevB.85.144402_2012,yang_intertwining_2022}. The resulting superconducting gap symmetry, for example, and the type of charge and spin density orders are qualitatively different, which is presumably attributed to the parameter choices as well as different numerical techniques. After the discovery of $A$V$_3$Sb$_5$, the two orbital model ($d_{yz}$ and $d_{zx}$) has also been adopted \cite{wu_nature_2021}. While its results carry invaluable new understanding especially for superconductivity, the difference from or inconsistency with other model studies is still notable. In order to have a better or hopefully complete understanding, it is important to identify the key characterizing factors for the given materials and to obtain realistic material parameters.

In this work, we perform the detailed electronic structure calculation of CsV$_3$Sb$_5$. We focus on the orbital character of the Van Hove singularity (VHS), its dependence on external strain and pressure, and the estimation of interaction strength. By defining the proper local coordinate transformation, we successfully construct the satisfactory orbital characterization of each band and VHS point, which provides a solid ground for model-based theoretical investigations. Remarkably, the out-of-plane Sb$^{\rm out}$-$p$ orbital is found to be crucial in the complete description of the three-VHS feature near the Fermi level ($E_F$). A systematic MLWF (maximally localized Wannier function) analysis clearly shows that this well-known VHS structure cannot be achieved  only with V-$d$ orbitals. Further, our cRPA (constrained random phase approximation) calculations reveal that Sb$^{\rm out}$-$p$ also plays a key role in the electronic screening process, thereby determining the effective correlation strengths. The realistic values of both on-site and inter-site interaction parameters are presented. By examining the effect of strain and pressure, we explore the possibility to control VHS through Sb$^{\rm out}$-$p$ states.



\vspace{3mm}\textbf{\\ Result and Discussion}$\;\;\;$

\begin{figure*}[t]
\includegraphics[width=0.95\textwidth]{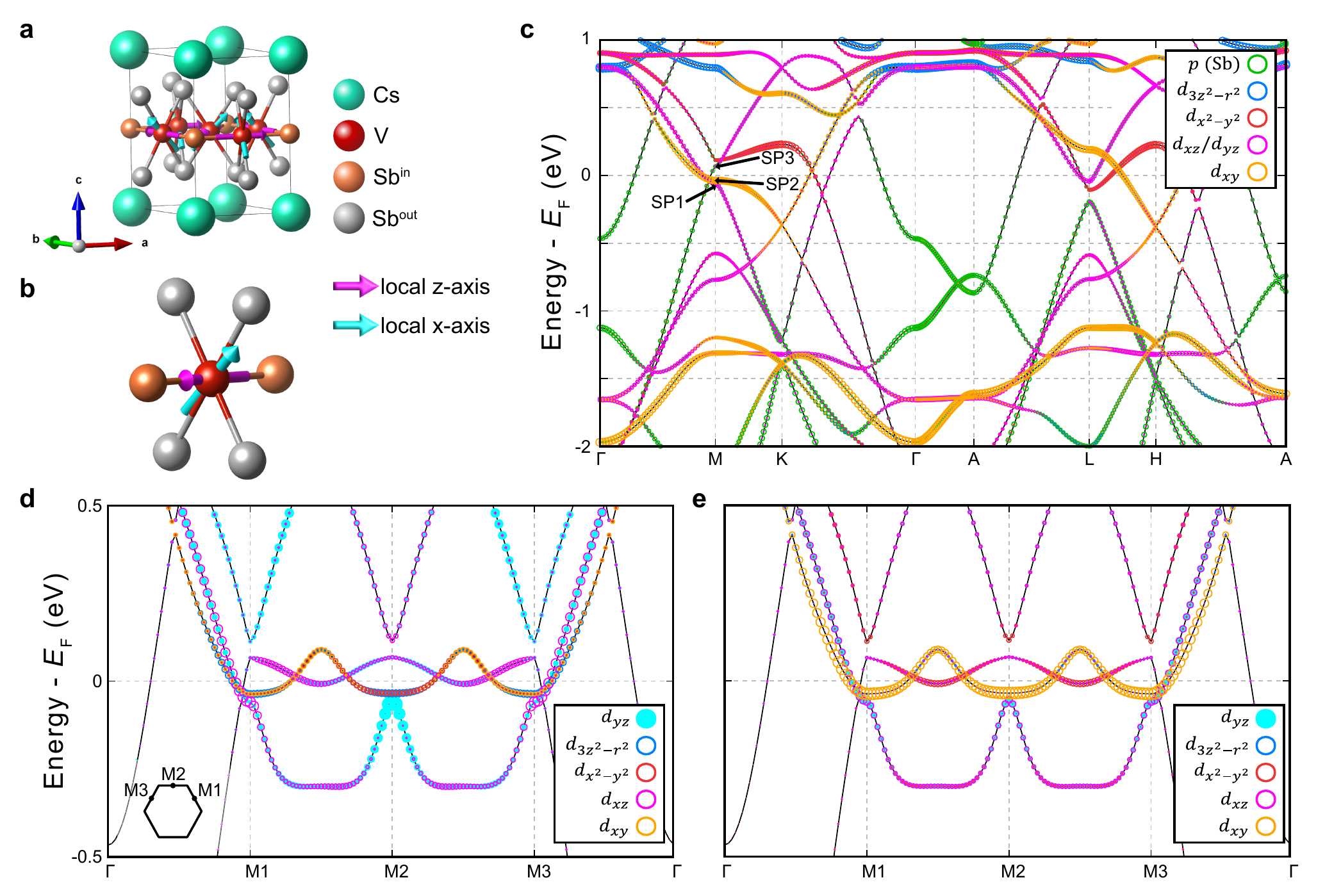}
\caption{\textbf{a} The crystal structure of CsV$_3$Sb$_5$. The cyan, brown, gray and red spheres represent Cs, Sb$^{\rm{in}}$, Sb$^{\rm{out}}$ and V atoms, respectively. \textbf{b} A local VSb$_6$ structure, where $x$ and $z$ axis are depicted by magenta and cyan arrow, respectively. \textbf{c} The orbital-projected electronic band structure of CsV$_3$Sb$_5$. The orbital projection is performed with the local coordinate as shown in \textbf{b}. The green, blue, red, magenta and orange color represent the Sb-$p$, V-$d_{\rm{3z^2-r^2}}$, V-$d_{\rm{x^2-y^2}}$, V-$d_{\rm{xz}}/d_{\rm{yz}}$ and V-$d_{\rm{xy}}$ characters, respectively, and the circle size shows the relative portion of them.  \textbf{d}, \textbf{e} The calculated band dispersion of CsV$_3$Sb$_5$ along the line connecting $\Gamma$ and three symmetric M points. The orbital projection was performed within the \textbf{d} global and \textbf{e} local axis coordinate. The filled cyan and open magenta circles represent the V-$d_{\rm{yz}}$ and V-$d_{\rm{xz}}$ character for \textbf{d} and \textbf{e}.}
\label{fig:local_global}
\end{figure*}


\noindent\textbf{Electronic structure and orbital characters.}
Figure~\ref{fig:local_global}\textbf{a} shows the crystal structure of CsV$_3$Sb$_5$. Separated by Cs layer,  V$_3$Sb$_5$ is composed of V$_3$Sb and two Sb layers. Two distinctive types of Sb, namely, Sb$^{\rm in}$ and Sb$^{\rm out}$, are located in the same plane with V and above/below the V-Sb$^{\rm in}$ layer, respectively. The V atoms make a Kagome network and are surrounded by six Sb atomic octahedron, leading to  $t_{\rm{2g}}$--$e_{\rm{g}}$ crystal field levels. The local structure of VSb$_6$ unit is shown in Fig.~\ref{fig:local_global}\textbf{b}. As discussed further below, it is important to take the proper local coordinate axis in analyzing V-$d$ energy levels because VSb$_6$ cages are not aligned along the same direction (see, e.g., Fig.~\ref{fig:local_global}\textbf{a}). Since the bond distance from V to Sb$^{\rm in}$ is different from that to Sb$^{\rm out}$, we take the local $z$-axis along the V--Sb$^{\rm in}$ line (the magenta arrow in Fig.~\ref{fig:local_global}\textbf{b}) which makes the $x$ and $y$ direction equivalent (the cyan arrow in Fig.~\ref{fig:local_global}\textbf{b}). Note that the local $z$-direction is perpendicular to the global c-axis of the unit cell; see Fig.~\ref{fig:local_global}\textbf{a}. This choice is also different from that of a previous study \cite{gu_gapless_2021_singleband_MLWF_model}.

Figure~\ref{fig:local_global}\textbf{c} shows the orbital projected band structure. Three saddle points (SPs) are clearly identified near $E_F$ at the M point. These VHSs are the key to understand the intertwining orders in this family of materials. For convenience, we name them SP1, SP2 and SP3 in the increasing order of energy. In Figs.~\ref{fig:local_global}\textbf{d} and \textbf{e}, the energy bands are recasted to highlight their orbital characters around three M points. It is noted that, only by taking the proper local coordinate axis, the orbital characters of all SPs become symmetric at the M.

It is important to note that SP2 is well described by single-orbital character of $d_{\rm{xy}}$. Its portion is $\sim 70\%$ at all three M points (Fig.~\ref{fig:local_global}\textbf{e}). It is not the case for the global-axis projection (Fig.~\ref{fig:local_global}\textbf{d}) in which SP2 has the mixed-orbital character and the composition is different at three M points. This SP2 VHS is important to understand the competing phases and therefore is taken as the basis for many one-orbital model studies \cite{feng_chiral_2021,feng_lowenergy_PhysRevB.104.165136,denner_analysis_2021,lin_complex_PhysRevB.104.045122_2021,park_electronic_2021,gu_gapless_2021_singleband_MLWF_model,jiang_unconventional_2021}. It is also this band dispersion for which the higher-order nature of the VHS was highlighted \cite{kang_twofold_2021,hu_rich_2021}. Thus the dominant single-orbital character identified by our local coordinate projection strongly supports this line of one-orbital (or three-band in the unit cell) tight-binding (TB) model studies and the patch models \cite{feng_chiral_2021,feng_lowenergy_PhysRevB.104.165136,denner_analysis_2021,lin_complex_PhysRevB.104.045122_2021,park_electronic_2021,gu_gapless_2021_singleband_MLWF_model,wang_competing_PhysRevB.87.115135_2013,Kiesel_unconvenitonal_PhysRevLett.110.126405_2013,Kiesel_sublattice_PhysRevB.86.121105_2012,yu_chiral_superconducting_PhysRevB.85.144402_2012,jiang_unconventional_2021,yang_intertwining_2022}, providing a solid ground for such theoretical works.

Another notable model approach takes two orbitals of $d_{yz}/d_{xz}$. For example, it was adopted for a recent RPA study to elucidate the unconventional superconductivity in this system \cite{wu_nature_2021}. Also in the ARPES analysis, the stabilization of CDW and the corresponding spectrum modulation have been discussed based on the two-orbital model \cite{kang_twofold_2021}. Two-orbital TB Hamiltonian  supports the two VHSs, corresponding to SP1 and SP3 in our density functional theory (DFT) band, and the higher-lying SP3 is known to play the key role in Fermi surface nesting and superconductivity \cite{wu_nature_2021,kang_twofold_2021}. Unraveling the detailed electronic nature of SP3 is therefore of crucial importance.

Fig.~\ref{fig:local_global}\textbf{e} shows that the main V-$d$ character for both SP1 and SP3 is in fact $d_{yz}/d_{xz}$. As projected onto the proper local coordinate, both orbitals equally participate in the VHSs and the related bands. We also note that SP3 is mixed-type with two-sublattice character as discussed in the literature \cite{denner_analysis_2021,wu_nature_2021,kang_twofold_2021}. An important and previously unnoticed feature for SP3 is that V-$d$ is not the major component at SP3. It is in contrast to the cases of SP1 and SP2 which are composed of $\sim 80$ and $\sim 70\%$ of V-$d$ character, respectively. For SP3, on the other hand, V-$d$ portion is only $\sim 32\%$ and the major contribution comes from Sb$^{\rm out}$-$p$ which reaches up to $\sim 50\%$ while Sb$^{\rm in}$-$p$ contribution is found to be negligible. In the following, we further highlight the critical role of Sb$^{\rm out}$-$p$ orbital particularly for the formation of VHS and the screening effect on interaction strength.

\begin{figure}[t]
\includegraphics[width=0.45\textwidth]{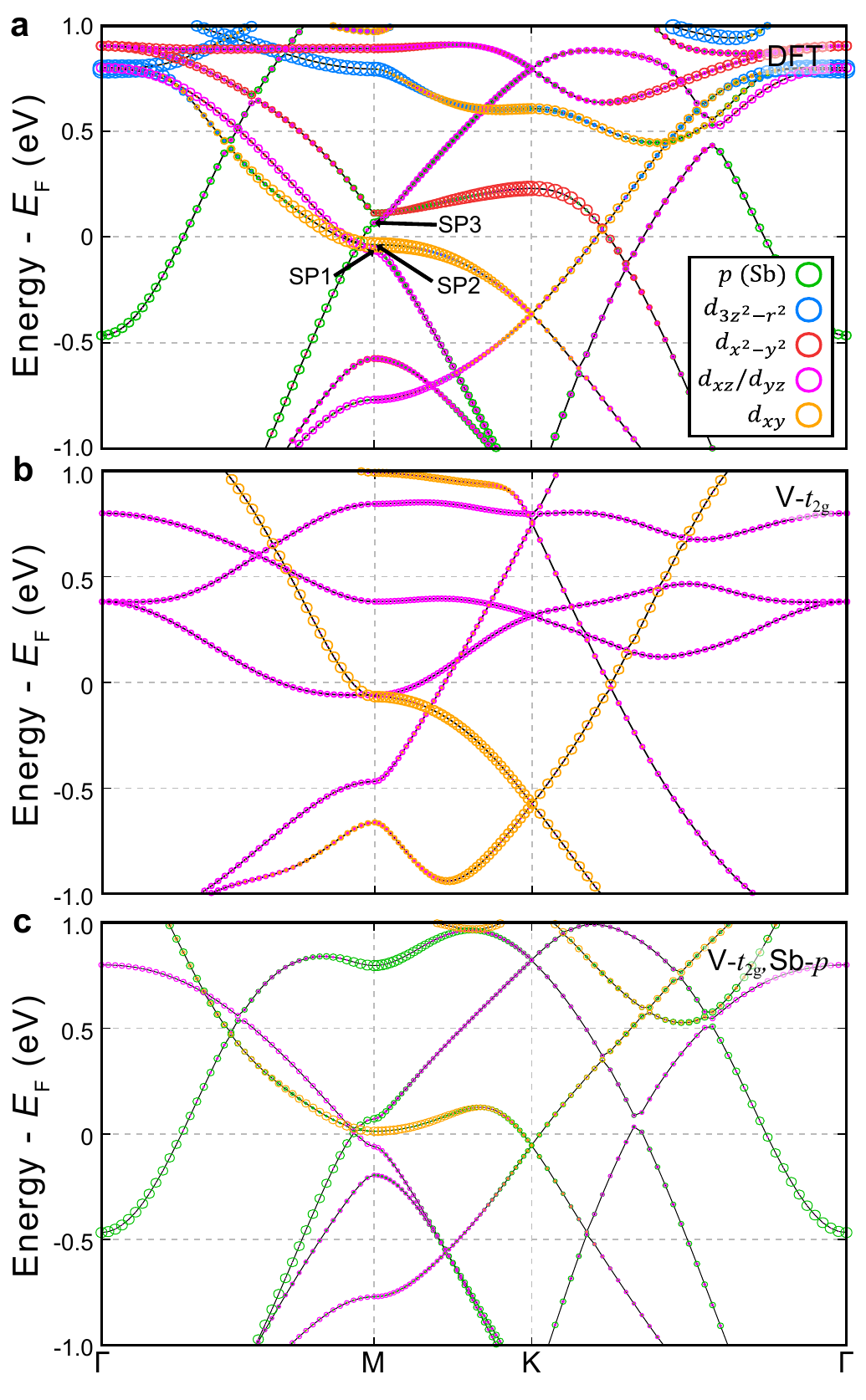}
\caption{\textbf{a} The electronic band structure computed by DFT. \textbf{b}, \textbf{c} The TB band structure as constructed by MLWF method. The TB model contains \textbf{b} V-$t_{\rm{2g}}$ orbital and \textbf{c} V-$t_{\rm{2g}}$ + Sb-$p$. The green, blue, red, magenta and orange circles represent the Sb-$p$, V-$d_{\rm{3z^2-r^2}}$, V-$d_{\rm{x^2-y^2}}$, V-$d_{\rm{xz}}/d_{\rm{yz}}$ and V-$d_{\rm{xy}}$ character, respectively. The size of circles depicts their portions at given points.}
\label{fig:TBmodel}
\end{figure}

\label{sec:parameteriation}
{\noindent\textbf{VHS formation: tight-binding analysis.}}
To understand the electronic details in the formation of VHSs, we perform a systematic TB analysis. Based on the standard MLWF technique \cite{souza_maximally_2001,marzari_maximally_1997}, we first construct a $t_{\rm{2g}}$-only (for each of three V atoms in the unit cell) TB model whose result is presented in Fig.~\ref{fig:TBmodel}\textbf{b}. Compared with the full band DFT result shown in Fig.~\ref{fig:TBmodel}\textbf{a}, it is noted that the main feature of $d_{xy}$ band and its near-$E_F$ SP nature at the M point are well reproduced although its dispersion is less flat than the DFT counterpart. Simultaneously, it is obvious that the $t_{\rm{2g}}$-only TB Hamiltonian cannot reproduce SP1 and SP3 VHSs: The $d_{xz/yz}$-dominant band shape (magenta colored) is largely different from the one in DFT, forming a bare parabolic minimum (rather than a SP) around $E_F$. Overall, the $t_{\rm 2g}$-only TB model only reproduces SP2 VHS with the noticeably enhanced single-orbital character. With this limitation, it can serve as a good effective model to focus on the VHS point with $d_{xy}$ character \cite{denner_analysis_2021,feng_chiral_2021,lin_complex_PhysRevB.104.045122_2021,wang_competing_PhysRevB.87.115135_2013,Kiesel_sublattice_PhysRevB.86.121105_2012,Kiesel_unconvenitonal_PhysRevLett.110.126405_2013,yu_chiral_superconducting_PhysRevB.85.144402_2012,park_electronic_2021,gu_gapless_2021_singleband_MLWF_model,feng_lowenergy_PhysRevB.104.165136}.

Notably, extending $t_{2g}$ manifold to the full V-$3d$ orbitals does not make a significant improvement in regard to VHS formation. Our result of the five-orbital TB model is presented in Supplementary Section 1. The main changes by having additional  $e_g$ orbitals is the extra bands appearing at higher energy above +0.5 eV, which are in fact mostly of $e_g$ character (see Supplementary Fig.2\textbf{a}). Regarding the VHS feature, only slight changes are noticed at the M point in the $d_{xz,yz}$ dispersion and the location of lower-lying VHS (at $\sim -0.5$ eV). SP1 and SP3 VHSs are still absent in this five-orbital model result. We note that not only three-orbital $t_{2g}$ but also five  V-$d$ orbital model cannot capture the three VHS features in this material.

Remarkably, the Sb$^{\rm out}$-$p$ orbitals play the key role in the formation of two more VHSs. Fig.~\ref{fig:TBmodel}\textbf{c} shows our TB result of V-$t_{\rm 2g}$ + Sb-$p$ model. Not only SP2 ($d_{xy}$-dominant; yellow colored), but also SP1 (mainly $d_{xz/yz}$-character; magenta colored) and SP3 (mixed $d_{xz/yz}$-character with Sb-$p$) are clearly identified although $d_{xy}$ band  gets deformed near $E_F$ away from the full band DFT result. In fact, SP3 itself has a sizable portion of Sb-$p$ character. Importantly, it is Sb$^{\rm out}$-$p$ orbitals that participate in the SP formation whereas Sb$^{\rm in}$-$p$ portion is negligible. As discussed in previous studies \cite{tsirlin_role_2021,labollita_tuning_2021} the main contribution of Sb$^{\rm in}$-$p$ in the low-energy physics is found in the $\Gamma$ pocket (see Fig.~\ref{fig:TBmodel}\textbf{a} and \textbf{c}).

Our new finding provides useful insight in understanding the formation of low-energy VHS and the intertwining orders in this material. As multiple VHS features are largely affected by Sb$^{\rm out}$, the energy scale of CDW and other related phases can possibly be changed and controlled by varying the type and/or the height of group V elements. It may also be related to the recently observed three-dimensional structure of CDW pattern \cite{liang_three-dimensional_2021,li_observation_2021,song_orbital_2021,ortiz_fermi_2021,ratcliff_coherent_2021,kang_microscopic_2022,hu_coexistence_2022}.

\label{sec:cRPA}

\noindent\textbf{Interaction strength: cRPA calculations.}
Electronic interaction is important information for understanding phases in this Kagome system. Previous theoretical studies revealed that several different phases can be stabilized by varying interaction parameters. Since the suggested phase diagrams are markedly different from one another depending on both on-site ($U$) and inter-site correlation ($V$) strengths as well as the details of model construction and computation \cite{wang_competing_PhysRevB.87.115135_2013,Kiesel_sublattice_PhysRevB.86.121105_2012,Kiesel_unconvenitonal_PhysRevLett.110.126405_2013,wu_nature_2021,denner_analysis_2021},
it is crucial to estimate the realistic values of interaction strengths. One of the most advanced `first-principles' schemes, namely cRPA, computes the screened Coulomb interactions by systematically removing the partial screenings which are not relevant to a given effective model \cite{aryasetiawan_frequency-dependent_2004,aryasetiawan_calculations_2006}. For Hubbard-type models with a given correlated subspace, one can therefore calculate the corresponding interaction parameters as wanted. For example, removing the screening effect among V-$d$ and Sb-$p$ yields the value for `$dp$ model'. One can also compute the interaction parameters corresponding to `$d$-band only' Hubbard model by removing only V-$d$ part (in cRPA literature, called `$d-dp$ model' \cite{sakuma_first-principles_2013,vaugier_hubbard_2012}).

\begin{table}[t]
\centering
\begin{tabular}{| c | c | c | c |}
 \hline
  Removed Screening & $U (F_0)$ (eV) & $J$ (eV) & $V$ (eV) \\
 \hline\hline
  All (Bare) & 15.81 & 0.78 & 3.50 \\ 
 \hline
  V-$d$, Sb-$p$ & 5.33 & 0.71 & 1.56 \\
 \hline
  V-$d$, Sb$^{\rm out}$-$p$ & 2.82 & 0.68 & 0.57 \\
 \hline
  V-$d$, Sb$^{\rm in}$-$p$ & 0.97 & 0.62 & 0.10 \\
 \hline
  V-$d$ & 0.87 & 0.61 & 0.10 \\
 \hline
\end{tabular}
\caption{The calculated on-site and inter-site interaction parameters for Hubbard-type model containing correlated V-$d$ orbital states. A systematic cRPA investigation provides the parameters for the given set of removed screenings. It is noted that, as more screenings get removed, greater interaction strengths are obtained as expected.
	For Hubbard $U$ and Hund $J$, we follow the definitions as described in Ref. \citenum{anisimov_density-functional_1993,anisimov_first-principles_1997,vaugier_hubbard_2012} for example.}
\label{table:cRPA}
\end{table}

Table~\ref{table:cRPA} presents our cRPA results of on-site interaction parameters Hubbard $U$ and Hund $J$ together with inter-site $V$. As expected, the more screenings get removed, the larger correlation parameters are obtained \cite{sakuma_first-principles_2013,vaugier_hubbard_2012}. While the `bare' interactions (corresponding to the atomic values) are quite large ($U=$15.81 and $V=$ 3.50 eV), the gradual inclusion of screening effects provide a reasonable estimation of interactions. When the screening effect of both V-$d$ and Sb-$p$ orbitals are removed (i.e. when `$dp$ model' is considered), $U$, $J$ and $V$ are found to be 5.33, 0.71, and 1.56 eV, respectively, in our VASP-cRPA calculation. For more details of computations, see, Method. Our `$dp$ model' results of $U=$5.3 and $J=$ 0.71 eV are quite close to the values used in a previous dynamical mean-field theory (DMFT) study, supporting their parameter choice \cite{zhao_electronic_2021}.

The values from `$d-dp$ model' are markedly different. By further excluding Sb-$p$ states, $U$ and $V$ is significantly reduced to 0.87 and 0.10 eV, respectively. While this feature of reduced interactions is a general trend of `$d-dp$ model' vs `$dp$ model' \cite{sakuma_first-principles_2013}, it provides a realistic set of interaction parameters for widely used `$d$-band only' Hubbard model.

Another remarkable and quite unexpected role of Sb$^{\rm out}$-$p$ states is found in determining interaction strengths. To examine the role of Sb$^{\rm out}$-$p$ state in the screening process, we perform cRPA calculations by separating Sb states into Sb$^{\rm out}$ and  Sb$^{\rm in}$, and the results are presented in Table~\ref{table:cRPA}; see the 4th and 5th row. By comparing the result of (V-$d$, Sb-$p$) with that of (V-$d$, Sb$^{\rm out}$-$p$) and  (V-$d$, Sb$^{\rm in}$-$p$), one can see that Sb$^{\rm out}$-$p$ states play the key role in determining correlation strength. The inclusion/exclusion of Sb$^{\rm out}$-$p$ most significantly affect the parameters. By including only Sb$^{\rm out}$-$p$ screening (excluding Sb$^{\rm in}$-$p$ screening) in the `$dp$ model', we obtain $U=$0.97 eV which is quite close to 0.87 eV of `full' $d-dp$ model result (see Method for more details). Hence, most of the screening effect in `$d-dp$' model comes from Sb$^{\rm out}$-$p$ states. It highlights the important role of Sb$^{\rm out}$ in determining the correlation physics of this material.

Here, we recall that Sb$^{\rm out}$-$p$ state is important in forming the correct VHS features around $E_F$ as discussed above. It indicates that both VHS physics and the correlation strength can be engineered by controlling Sb$^{\rm out}$. In iron-based superconductors, the relation of pnictogen height to correlation and magnetic coupling has been highlighted  \cite{iron5_chen_iron-based_2014,iron4_kuroki_pnictogen_2009,iron3_mizuguchi_anion_2010,iron2_okabe_pressure-induced_2010,iron1_moon_chalcogen-height_2010,Yin_Pickett_PRL2008_PhysRevLett.101.047001}. Useful insights may therefore be obtained by examining the change of electronic properties in response to pressure and strain.


\begin{figure*}[ht]
\includegraphics[width=0.95\textwidth]{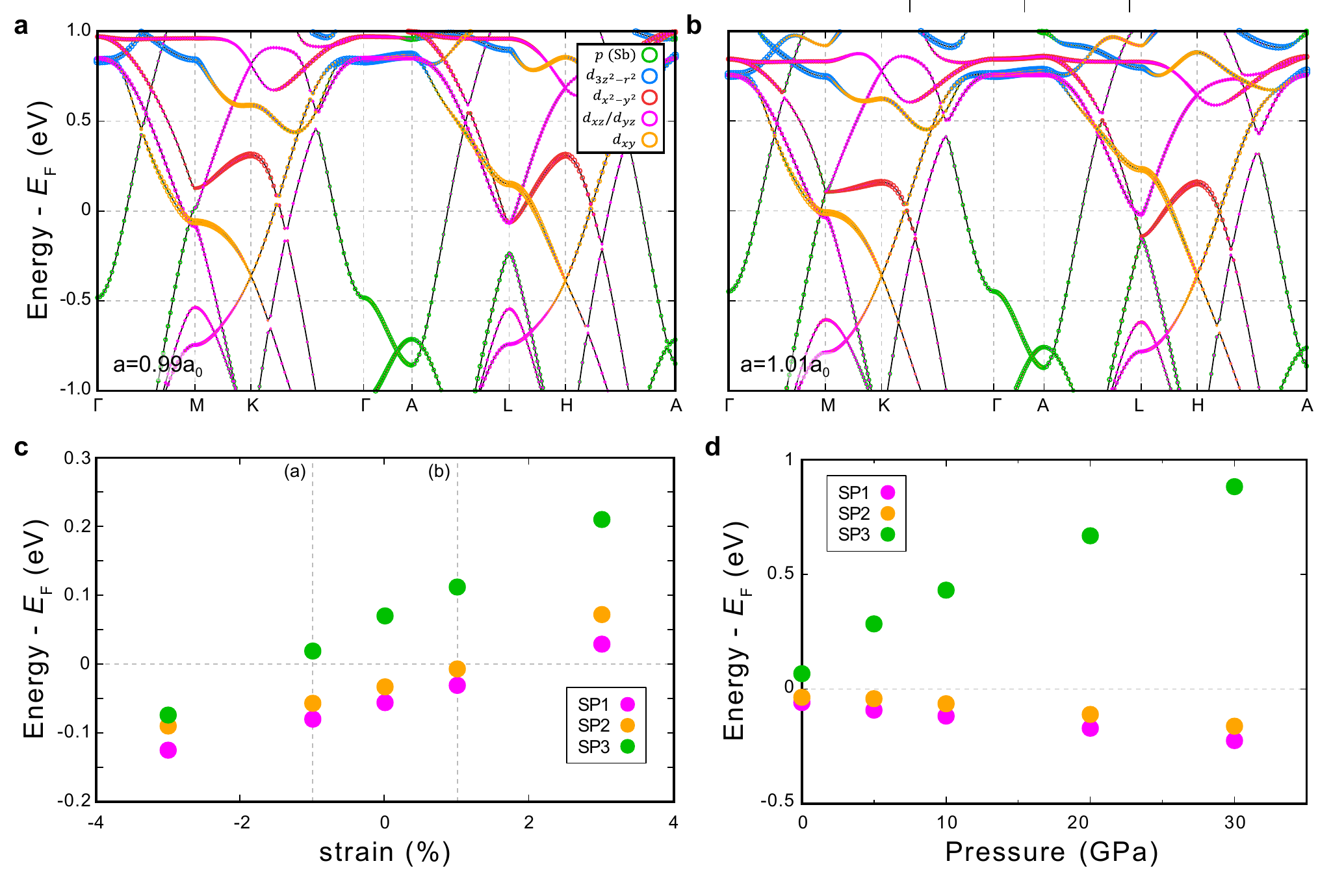}
\caption{\label{fig:strain} \textbf{a}, \textbf{b} The orbital-projected electronic band of CsV$_3$Sb$_5$ for 1$\%$ of \textbf{a} compressive and \textbf{b} tensile bi-axial strain. The green, blue, red, magenta and orange color represent the Sb-$p$, V-$d_{\rm{3z^2-r^2}}$, V-$d_{\rm{x^2-y^2}}$, V-$d_{\rm{xz}}/d_{\rm{yz}}$ and V-$d_{\rm{xy}}$ characters, respectively. \textbf{c}, \textbf{d} The calculated energy position of three SPs by varying \textbf{c} strain and \textbf{d} pressure. The vertical dashed lines in \textbf{c} show the data correspond to \textbf{a} and \textbf{b}.}
\end{figure*}

\label{sec:strain}
{\noindent\textbf{The effect of {bi-axial} strain and pressure.}}
Both pressure and (uni-axial) strain are proven to be the relevant external stimuli for this material family \cite{song_competing_2021,song_competition_2021,qian_revealing_2021_strain,chen_double_2021,yu_unusual_2021,zhang_pressure-induced_2021,du_pressure-induced_2021}. Before discussing the change and control of VHSs, let us first check if they are still well defined and the same orbital characters are maintained in an experimentally accessible range of strain. Figure~\ref{fig:strain}\textbf{a} and \textbf{b} show the calculated band structures under 1\% of compressive and tensile strain, respectively. The overall dispersion, as well as the orbital characters, are retained, and therefore one can easily identify each SPs and monitor their evolution. We find that it is also the case for hydrostatic pressure (not shown). For more information including the change of higher-order feature of SP2 band, see Supplementary Section 2.

The evolution of three VHS positions as a function bi-axial strain is presented in Fig.~\ref{fig:strain}\textbf{c}. An obvious trend is clearly observed: Their energy positions all gradually increase as more tensile strain is applied (the in-plane lattice parameter increases). Our result suggests an interesting possibility; that is, the relative position of VHS points can be systematically controlled by strain. At zero strain, for example, SP2 is closest to $E_F$ (corresponding to zero energy reference in the figure), and it gets even closer at 1\% tensile strain. At $-$1\% of (compressive) strain, on the other hand, SP3 is the closest VHS to $E_F$. Our results imply that the strain dependent experiment can provide important information about the role of VHSs and/or their cooperations/competitions in this intriguing system.

The effect of pressure is fairly different in this regard. Figure~\ref{fig:strain}\textbf{d} shows the calculated SP eigenvalues as a function of hydrostatic pressure. An increasing trend of the SP3 position is clear, which is distinctive from the decreasing SP1 and SP2 positions \cite{tsirlin_role_2021,labollita_tuning_2021}. All three VHS positions move away from $E_F$ and therefore the order of their relative distances to $E_F$ is not changed.

SP3 moving toward higher energy by pressure is in contrast to the case of compressive strain in which the reduced in-plane lattice constant makes its position lower in energy. This different behavior can be understood by considering the out-of-plane lattice parameter which is reduced by applying pressure but enlarged by compressive strain. Here, we once again notice the role of Sb$^{\rm out}$-$p$ which makes significant contribution to SP3 VHS while SP1 and SP2 are mainly composed of V-$d$ orbitals. According to recent experiments, SP3 could mainly be responsible for Fermi surface nesting \cite{kang_twofold_2021}. Also, $T_{\rm CDW}$ decreases as a function of out-of-plane lattice constant in both strain and pressure experiments \cite{qian_revealing_2021_strain}. Further study focusing on SP3 might be able to shed new light on  many issues in this system.

\label{sec:summary}
\vspace{3mm}\textbf{\\ Summary}$\;\;\;$

To summarize, we performed the DFT calculations on CsV$_3$Sb$_5$ and pointed out the importance of Sb$^{\rm out}$-$p$ orbitals. Near $E_F$, three VHSs (V-$d_{xy}$ dominant, V-$d_{yz}/d_{xz}$ dominant, and V-$d_{yz}/d_{xz}$+Sb-$p$) are found. By TB model generation, we found that except $d_{xy}$-dominant VHS, other two VHSs cannot be formed without the consideration of Sb$^{\rm{out}}$-$p$ orbitals. This Sb$^{\rm{out}}$ atoms are responsible for not only the formation of VHSs, but also provide significant screening effect. Considering the effect of Sb$^{\rm{out}}$ for the VHS formation and correlation, we proposed to control the VHS features by the strain and pressure, which will change the height of Sb$^{\rm{out}}$. For example, the energy level of SP3 is connected to out-of-plane lattice constant which can be changed by the strain or pressure. Elucidating the relationship between energy level of these VHSs and CDW instability would be an outstanding topic of future study.



\vspace{3mm}\textbf{\\ Methods}$\;\;\;$ 

\noindent\textbf{First-principles calculation.}
We carried out first-principles density functional theory (DFT) calculations. For the structural optimizations, we mainly used `Vienna Ab initio Simulation Package (VASP)' based on the projector augmented-wave pseudopotential \cite{kresse_ab_1993,kresse_efficiency_1996} within PBE generalized gradient approximation (GGA) \cite{perdew_generalized_1996}. Both lattice parameters and internal atomic coordinates were optimized with a force criterion of 0.001 eV/\AA. The $8\times8\times4$ k-points and the 500 eV energy cutoff were adopted. For simulating the strained conditions, the out-of-plane lattice constants and the internal coordinates were calculated at the fixed lateral lattice values. The van der Waals interaction has been taken into account within so-called `DFT-D3' functional scheme \cite{grimme_consistent_2010}. For the electronic structure analysis, we mainly used  `OPENMX' software package based on linear combination of pseudo-atomic-orbital basis \cite{ozaki_PhysRevB.67.155108} and within PBE-GGA exchange-correlation functional \cite{perdew_generalized_1996}. $12\times12\times6$ k-grids and 400 Ry energy cutoff were used. Maxiamlly localized Wannier function (MLWF) method \cite{marzari_maximally_1997,souza_maximally_2001} has been used to construct tight-binding (TB) models.
For further orbital-dependent analysis we also used `DFTforge' code (which is a part of our `Jx' code) \cite{yoon_jx_2020}.

To calculate the interaction parameters, constrained radom phase approximation (cRPA) calculation was conducted \cite{aryasetiawan_frequency-dependent_2004,aryasetiawan_calculations_2006,sasioglu_effective_2011}. For this purpose we used `VASP' (with `Wannier90' \cite{Pizzi2020})  with $4\times4\times2$ k-mesh. Each orbital space of V-$d$ and Sb-$p$ is defined through the corresponding MLWFs. The on-site interaction parameters $U$ and $J$ are defined following Ref.~\citenum{anisimov_density-functional_1993,anisimov_first-principles_1997,vaugier_hubbard_2012}. The nearest-neighboring inter-site interaction $V$ is defined as the average value:
\begin{equation}
    V=\frac{1}{N}\sum_{i\neq j,\alpha,\beta}W_{i\alpha,i\alpha,j\beta,j\beta}
    \label{eq:inter-site}
\end{equation}
where $W$ is the interaction matrix with the sublattice indices $i,j$, the orbital $\alpha,\beta$, and $N$ refers to the number of inter-site interaction terms. A part of the results are double-checked with `ECALJ' code \cite{kotani_ecalj}.

As mentioned in the main text, cRPA procedure provides the systematic estimation of interaction parameters by properly defining so-called `target space' and `rest space'. The effective screened interactions are computed from bare-Coulomb interaction ($\mathcal{V}$) by considering the rest space polarization ($P_{\rm R}$) as follow \cite{aryasetiawan_frequency-dependent_2004,aryasetiawan_calculations_2006}: 
\begin{equation}
    U=[1-\mathcal{V}P_{\rm R}]^{-1}\mathcal{V}.
\end{equation}
To obtain $\mathcal{V}$ and $P_{\rm R}$, we need to define target space. One conventional approach is using MLWF to define the target and rest spaces \cite{sasioglu_effective_2011,sakuma_first-principles_2013,nomura_ab_2012,jang_direct_2016,vaugier_hubbard_2012,miyake_cRPA1_PhysRevB.77.085122,miyake_cRPA2_PhysRevB.80.155134}. Supplementary Figure~1\textbf{a} and \textbf{b} (in Supplementary Information) show the choice of target and rest space corresponding to so-called `$dp$' and `$d-dp$' model in cRPA literacture. The target space polarization ($P_T$; red arrows) includes both V-$d$ and Sb-$p$ orbital space in `$dp$' model, and therefore the transitions within V-$d$, Sb-$p$ and V-$d$ + Sb-$p$ are taken into account for $P_{\rm T}$. On the other hand, so-called $d-dp$ model only takes V-$d$ space as its target space.\\[5mm]

\label{sec: Acknowledgements}
\vspace{3mm}\textbf{\\ Acknowledgements}$\;\;\;$

M.Y.J. and M.J.H. are supported by the National Research Foundation of Korea (NRF) grant funded by the Korea government (MSIT) (No. 2021R1A2C1009303 and No. 2018M3D1A1058754).
M.Y.J. and M.J.H. are supported by the KAIST Grand Challenge 30 Project (KC30) in 2021 funded by the Ministry of Science and ICT of Korea and KAIST (N11210105).
H.J.Y., H.S.K. and S.B.L. are supported by NRF Grant (No. 2020R1A4A3079707, No. 2021R1A2C1093060). 
Y.B.K. is supported by the NSERC of Canada and the Center for Quantum Materials at the University of Toronto.

\vspace{3mm}\textbf{\\ Author contributions}$\;\;$

Y.B.K., S.B.L. and M.J.H. conceived the work. M.Y.J. and M.J.H. performed theoretical calculations.  All authors discussed the results and wrote the manuscript.

\vspace{3mm}\textbf{\\	Additional information}$\;$

Correspondence to Yong Baek Kim, SungBin Lee and Myung Joon Han

\vspace{3mm}\textbf{\\	Data availability}$\;$

 All relevant data are available from the corresponding author upon reasonable request.

\vspace{3mm}\textbf{\\	Competing financial interests}$\;\;\;$

The authors declare that they have no competing financial interests.\\[3mm]

\bibliography{V_Kagome_DFT}

\end{document}